\documentclass[a4paper]{article}
\usepackage[margin=25mm]{geometry}
\usepackage{amsfonts}
\usepackage{amssymb}
\usepackage{graphicx}
\usepackage{svg}
\usepackage{amsmath}
\usepackage{caption}
\usepackage[labelfont={small,bf}]{caption}
\usepackage[font={small}]{caption}
\usepackage{subcaption}
\usepackage{dsfont}
\usepackage{float}

\usepackage{hyperref}
\newcommand{\red}[1]{{#1}}
\newcommand{\partialder}[1]{\frac{\partial{#1}}{\partial{t}}}

\title{
From Toggle to Tuning: Controlling Turing Patterns in Gene Circuits}
\author{{Antonio Matas-Gil}$^{1,2\ddagger}$, {Robert G. Endres$^{2 \dagger}$}}

\date{\small $^{1}$Division of Infection and Immunity \& Institute for the Physics of Living Systems, University College London, London WC1E 6BT, United Kingdom \\
$^{2}$Department of Life Sciences \& Centre for Integrative Systems Biology and Bioinformatics, Imperial College London, London SW7 2AZ, United Kingdom
\\\vspace{0.5cm}$^\dagger$r.endres@imperial.ac.uk,  $^\ddagger$a.matasgil@ucl.ac.uk
}

\begin{document}
\maketitle

\section*{Abstract}
Controlling spatial patterns in synthetic biological systems remains challenging due to poor parameter robustness and limited experimental tunability. We introduce two complementary mechanisms—the pattern switch and the pattern dial—to systematically control Turing pattern formation in gene circuits. The switch toggles pattern onset via a single parameter, while the dial enables transitions between distinct pattern types using weakly nonlinear amplitude equations. Analyzing network size reveals a key trade-off: small networks are easier to control but less robust, while larger networks gain robustness at the cost of tunability—suggesting a sweet spot for both evolvability and designability. Our results offer practical design rules for engineering programmable patterns in living systems.\\

\section*{Introduction}
Patterns are ubiquitous in nature, and consequently pattern-formation mechanisms too. Pattern formation mechanisms are well known in physics such as different types of equilibrium and nonequilibrium phase separation, but in biology two are mainly discussed - Turing patterns \cite{Turing1952} and positional information \cite{Wolpert1969-wy}, although there are others e.g. motility-induced phase transitions and chemotaxis \cite{Martínez-Calvo_Wingreen_Datta_2023}. Positional information relies on a naturally occurring gradient and pre-defined thresholds, which allow cells to know their relative position to the source of the gradient \cite{Wolpert1969-wy}. In contrast, Turing patterns rely on diffusion and nonlinear interactions and feedbacks, causing a homogeneous steady state to become unstable due to noise, leading to a self-organized pattern of a specific wavelength. This lack of physical impositions has led to a significant amount of research into Turing patterns as a possible mechanism in the formation of fingerprints \cite{fingerprintturing}, digits \cite{digitPatterning} and hair follicles \cite{hairPatterning} in mice, skin and scale patterns e.g. in zebrafish \cite{fishPatterning}, as well as patchy vegetation in ecosystems \cite{manderMorphometricAnalysisVegetation2017, geHiddenOrderTuring2023}.

In many of these examples, the compounds involved in pattern formation seem to follow the Turing mechanism, particularly the short-range activation and long-range inhibition logic proposed by Gierer and Meinhardt \cite{giererMeinhardt}. However, the underlying biology of pattern formation is often far more complex and, as shown for digit formation in mice, may also involve additional morphogen gradients \cite{digitPatterning}, making it challenging to definitively attribute a given pattern to a Turing mechanism. Prepatterns, chemical gradients, entrainment by oscillators, and larger more complex networks of interacting molecules are known to strongly influence Turing patterns, and to make them more reproducible in biological contexts \cite{fingerprintturing, hairPatterning, stochasticTuring, shaberiOptimalNetworkSizes2025}.

Given the challenges of identifying Turing systems in natural patterning networks, some research has focused on reproducing Turing patterns in the lab. The first success came from chemical experiments, where the chlorite-iodide-malonic-acid reaction was found to undergo a Turing instability and lead to pattern formation \cite{chemTP1990}. Another focus is the synthetic engineering of Turing patterns in cells or simpler organisms like bacteria. Until recently, attempts to engineer such patterns had seen limited success. For instance in 2018, a stochastic model was used to describe random Turing-like patterns in bacterial populations, but although both experimental and simulation results were similar, neither produced a fully satisfactory regular pattern \cite{stochasticTuring}. The difficulty of engineering Turing patterns in bacteria was showcased in a study by Scholes {\it et al.} \cite{SCHOLES2019243} in 2019, where Turing networks with two and three nodes (or molecular species) were comprehensively analyzed, revealing that although around 61\% of the networks could theoretically produce Turing patterns, this occurred in less than 0.1\% of the biologically plausible parameter space. This finding shows that the robustness of Turing patterns to perturbations in parameter space is minimal.

Motivated by the latter study, bacteria were recently engineered by designing a genetic circuit based on the most robust Turing topology \cite{SCHOLES2019243} and subsequently grown in bacterial colonies, leading, indeed, to regular patterns \cite{martina2023}. Comparison of the numerical simulations and the experimentally produced Turing patterns agreed in different aspects, such as the wavelength, antiphase, and the behaviors of growth. Although this approach was more robust, the resulting patterns still lacked the expected consistency and repeatability. Generally, predictions and analyses of models are based on linearization around a steady state, requiring stability without diffusion but instability with diffusion. The stability is analyzed on the basis of the eigenvalues of the Jacobian matrices. As they become wavenumber-dependent, they are called dispersion relations. At the bifurcation point, when they change from a negative (stable) to a positive (unstable) sign, patterns can form. While a powerful tool, the dispersion relations do not predict the symmetry or type of pattern, e.g. stripes, spots, or labyrinths. This leads to a loss of predictability of such approaches.

\begin{figure*}[t!]
\includegraphics[width=\textwidth]{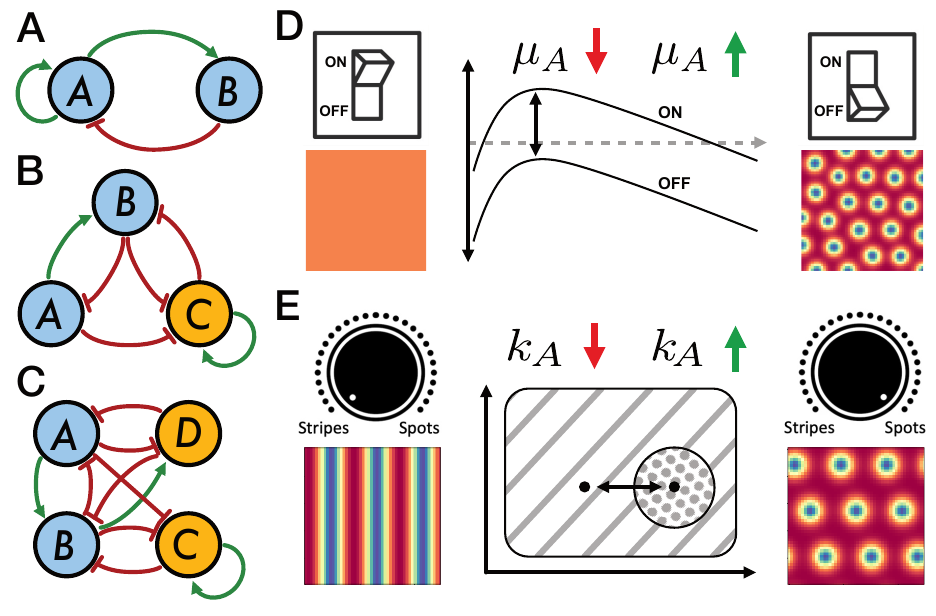}
\caption{\textbf{Controlling the dispersion relation for pattern formation in genetic networks.} \textbf{(A-C)} Representation of the genetic networks used in this work, corresponding to the 2-node (A), 3-node (B) and 4-node (C) networks, all capable of producing Turing patterns. \textbf{(D)} Pattern switch using the dispersion relation to turn on and off pattern formation. If the dispersion relation is below zero for all $k$, the system is stable and there is no pattern formation, corresponding to the switch being off. Through a parameter change, the dispersion relation can become positive, turning the switch for pattern formation on. Specifically for our initial parameter set, we found that $\mu_A$ can shift the dispersion relation, acting as a switch. \textbf{(E)} Pattern dial between different region of pattern stability. Weakly nonlinear analysis techniques is applied to investigate the stability of different patterns in parameter space, allowing us to find parameters that are capable of causing this transition. In this example, a low value of $k_A$ leads to stripes, while increasing it leads to spots.}
\label{fig1}
\end{figure*}

In this paper, we aim to bridge this gap by developing strong theoretical predictions that can be experimentally verified. Our goal is to derive conditions that confirm whether an observed experimental pattern is a genuine Turing pattern generated by \red{one of several genetic circuits (Fig. \ref{fig1} A-C)}, and to establish a framework for controlling patterns at various levels. We focus on two key approaches: the pattern ``switch'' and the pattern ``dial''. The pattern switch refers to the ability to toggle the formation of patterns on or off by adjusting specific parameters. We achieve this by minimizing the maximum value of the dispersion relation once a pattern has emerged, thereby controlling the onset of pattern formation. In contrast, the pattern dial focuses on altering the type of pattern produced. This is achieved by finding a set of parameters that allows a gradual transition from a spotted pattern to a striped one, akin to turning a dial. For this purpose, we use weakly nonlinear techniques and amplitude equations, applying them to a one-dimensional projection of the original Turing model onto the Swift-Hohenberg equation \cite{Swift_Hohenberg_1977}. This allows for a description of the stability of each pattern type in parameter space which we use to select parameters that allow transitions from regions where one pattern is stable to another region where a different pattern is stable. Both mechanisms, presented in Figs. \ref{fig1}D and E, are investigated in the context of realistic genetic network models, giving examples of how the our results can be used in an experimental setting. 

\section*{Results}
\subsection*{Pattern switch: controlling the onset of pattern formation}
 We begin by investigating toggling pattern formation on and off by modifying specific aspects of the dispersion relation for a given model and set of parameters. As a toy example, we consider the Schnakenberg model \cite{Schnakenberg1979}:
\begin{align}\label{schnakmodel}
\begin{split}
	\partialder{u}&=D_u\Delta u+ c_1 - c_2 u + c_3 u^2 v, \\ 
        \partialder{v} &= D_v\Delta v + c_4 - c_3 u^2 v,
 \end{split}
\end{align}
\red{where $u$ and $v$ are two-dimensional functions representing concentrations of two interacting substances, $D_{x}$ are the diffusion constants for $x=u,v$ and $c_1-c_4$ are rate constants.}
By applying the Fourier transform to the linearization around the steady state $(u_0,v_0)$ of this equation, we  arrive at the diffusion-dependent Jacobian:

\begin{equation}
\mathbf{J}_k = 
\begin{pmatrix}
-c_2+2 c_3  u_0 v_0 - D_u k^2 & c_3 u_0^2 \\
-2 c_3 u_0 v_0 & -c_3 u_0^2 - D_v k^2
\end{pmatrix}
\end{equation}
with $k=|{\mathbf k}|^2$ the wavenumber.
 The dispersion relation is defined as the maximum of the real parts of the eigenvalues of $\mathbf{J}_k$, which we can write as:

 \begin{equation}
\max{\Re( \lambda_{i}(k^2))} = \frac{-\text{Tr}(\mathbf{J}_k)+\Re\big(\sqrt{\text{Tr}(\mathbf{J}_k)^2-4D_u D_v \text{det}(\mathbf{J}_k)}\big)}{2 D_u D_v},
\end{equation}
where $\text{Tr}(\mathbf{J}_k)$ and $\text{det}(\mathbf{J}_k)$ denote the trace and determinant of $\mathbf{J}_k$, respectively. Pattern formation is dependent on the dispersion relation reaching a positive value at a given wavenumber: if it does, pattern formation is triggered; if it does not, no pattern formation occurs. If we denote $\lambda_{max}$ as the maximum of the dispersion relation, taking place at wavenumber $k_c^2$, such that $\max{\Re( \lambda_{i}(k^2))} = \lambda_{max}$, we can say that pattern formation occurs only if $\lambda_{max}>0$. By tuning this value, we can effectively create a switch that controls whether pattern formation occurs.
To find the best parameters that can serve as switches, we compute the gradient of the expression for the maximum with respect to all parameters and consider the largest component. Note that in this toy example, every step can be done analytically, since we can easily find closed-form expressions for $u_0$ and $v_0$, but this will not be the case for all models.

As an example of a simple pattern switch, we consider the following equation from motility-induced phase transitions, which naturally appear in our derivations later on \cite{Martínez-Calvo_Wingreen_Datta_2023}:

 \begin{equation}
\label{RDnewmodel}
\begin{split}
    \partialder{u} &= -D_u (\Delta+\kappa)^2 u +c_1 u + c_2 u v^2,\\
    \partialder{v}& = D_v \gamma \Delta v + c_3 v + c_4 v^2,
\end{split}
\end{equation}
with now the biharmonic operator $(\Delta+\kappa)^2$.
 We can easily find that the peak of the dispersion relation is determined by the parameter $c_1$ since the Jacobian matrix is diagonal, forming a perfectly simple parameter switch. Whenever $c_1>0$ pattern formation occurs.

\subsubsection*{Pattern switch by varying a single parameter}
 Now that we have introduced the idea of a parameter switch, we look into how to obtain it in practice. In particular, we obtain parameter switches for the 2-node network model, used to describe gene expression e.g. in synthetic circuits for pattern formation \cite{SCHOLES2019243, martina2023}:

 \begin{equation}\label{network2d}
\begin{split}
    \partialder{A} &= b_A+V_A\frac{1}{1+(k_A/A)^2}\frac{1}{1+(B/k_{BA})^2}-\mu_A A+D_A \Delta A ,\\
    \partialder{B} &= b_B+V_B\frac{1}{1+(k_{AB}/A)^2}-\mu_B B+D_B \Delta B,
\end{split}
\end{equation}
with parameters $b_X, V_X, k_X, \mu_X$ and $D_X$ describing respectively rates of basal expression, rates of maximal expression, concentration thresholds, degradation rates and diffusion constants for species $X=A,B$. As explained, we achieve control of pattern formation by investigating the effect of parameter changes in the dispersion relation. Specifically, our goal is to determine how different parameters influence $\lambda_{max}$ and to identify the optimal combination of parameters to achieve significant changes in this value. This value can be computed from the Jacobian $J$ of the linearized dynamics. Hence, not only do the parameters involved in $J$ play a role here, but also all other parameters of the model, since the Jacobian is taken around the steady state, which changes (albeit only slightly) if we change parameter values.


\begin{figure*}[t!]
\includegraphics[width=\textwidth]{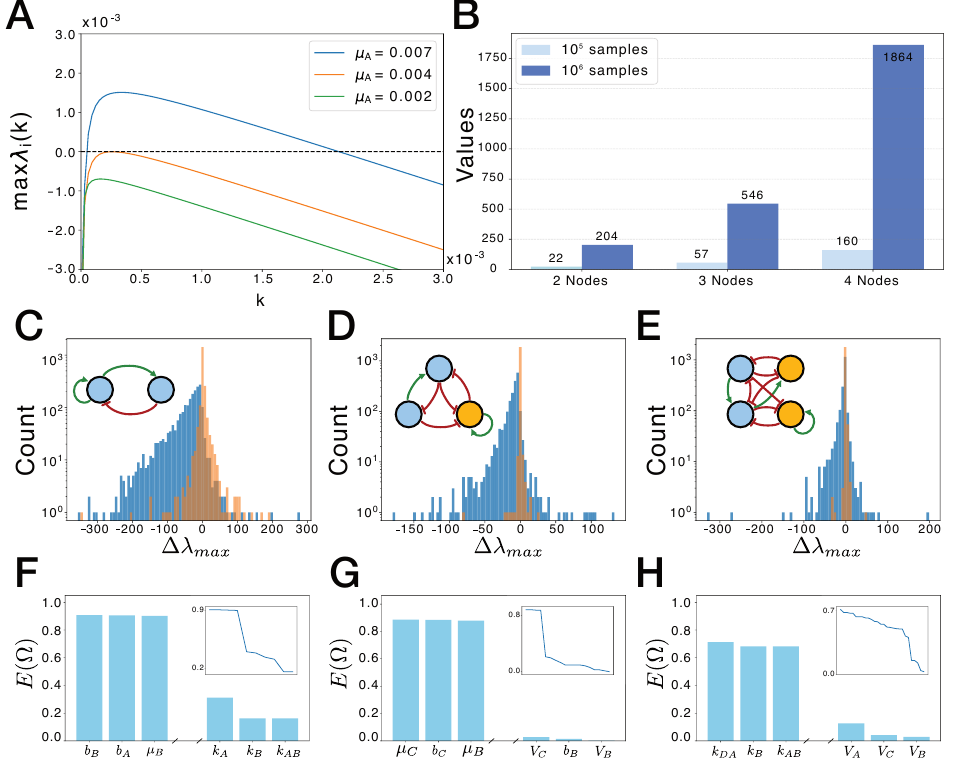}
\caption[\textbf{Control of the dispersion relation for pattern formation.}]{\textbf{Control of the dispersion relation for pattern formation.} \textbf{(A)} Dispersion relations for different values of the parameter $\mu_A$, identified as optimal for reducing the peak \red{for the initial parameter set}. When \red{$\mu_A = 0.002$}, the dispersion relation remains below zero, indicating that the system is stable with respect to diffusion and does not produce patterns. As $\mu_A$ increases, $\lambda_{max}$ approaches the critical value of 0 at approximately $\mu_A=0.004$. Beyond this point, the system becomes unstable to diffusion and is capable of forming patterns, in agreement with linear stability analysis. \red{A more complete plot showing the behavior of $\lambda_{max}$ as a function of both $\mu_a$ and $k$ is shown in Fig. S1C of the SI.} Parameters used: $D_A = 1, D_B = 1000, V_A = 21.544, V_B = 0.464, k_A = 21.544, k_{AB} = 21.544, k_{BA} = 0.464, b_A = 0.010, b_B = 0.010, \mu_A = 0.002, \mu_B = 0.010$. \textbf{(B)} Bar charts showing the number of Turing parameter sets identified for each model, with two different sample sizes: light blue represents the $10^5$ sample, and dark blue represents the $10^6$ sample. In the 2-node network, we found 22 Turing parameter sets in the $10^5$ sample and 204 in the $10^6$ sample, representing approximately 0.0002\% of the parameter space. For the 3-node network, we identified 57 and 548 Turing parameter sets in the respective samples (0.00055\%), and in the 4-node network, 160 and 1,864 Turing parameters were found, representing around 0.0017\% of the explored parameter space. These findings align with previous results \cite{SCHOLES2019243}, indicating that larger networks exhibit a greater abundance of Turing patterns in parameter space. \textbf{(C-E)}  Histograms of the highest (blue) and lowest (orange) ranked parameters based on the sign difference metric for the three models. The best parameters for the 2-node and 3-node networks, shown in (C) and (D), achieved a sign difference of approximately 0.9, indicating that most of the distribution lies on one side of the $x$-axis. In contrast, the best value for the 4-node network, shown in (E), is around 0.7, suggesting reduced controllability of the dispersion relation, and consequently, of pattern formation, as network complexity increases.\red{\textbf{(F-H)} Barplots of the sign difference score for the different parameters of each model in decreasing order. We can observe that for the 2- and 3-node models (F and G, respectively), the top parameters all have a very high sign difference score, and then experience a sudden drop. This drop is shown in the inner plots, which show the sign difference score in decreasing order for all parameters. In contrast, for the 4-node model (H), we observe a lower initial value and a continuous decrease, with a wider drop at the end. }}
\label{fig2}
\end{figure*}

 We began by computing these results analytically for a specific parameter set from Eq. \ref{network2d}, which we refer to as the 'initial parameter set', and is specified in the caption of Fig. \ref{fig2}A. Firstly, as shown in the toy model, we obtain an expression for $\lambda_{max}$ by finding the maximum of the dispersion relation with respect to $k^2$. Subsequently, we compute derivatives with respect to each parameter and numerically evaluate them to calculate the effect of small changes in each parameter on $\lambda_{max}$. This is similar to a derivative-based sensitivity analysis \cite{Renardy_Hult_Evans_Linderman_Kirschner_2019}. This approach does not consider the dependence of the Jacobian on the steady state, but we also computed the total derivative, which takes into account the changes in the steady state because of parameter changes. Nevertheless, this did not change the results obtained. 
 
 For the initial parameter set, our calculations showed that a single parameter modified $\lambda_{max}$ the most by several orders of magnitude compared to the rest: the degradation rate of the activating species, $\mu_A$. In Fig. \ref{fig2}A, we show the dispersion relation for different values of $\mu_A$. We observe that small changes in $\mu_A$ are capable of switching $\lambda_{max}$ from positive to negative, thereby switching pattern formation on and off. Specifically, we observe that when $\mu_A$ is close to $0.004$, $\lambda_{max}$ nears the zero critical value. Hence, if $\mu_A>0.004$ we satisfy the conditions from linear stability analysis for pattern formation, while when $\mu_A<0.004$, we cannot have pattern formation.
 
This pattern switch is parameter-specific, meaning that if we change the initial parameters of the model, the results will likely also change. Hence, instead of having a single parameter being the best switch, it might be a different combination of parameters that achieves the best possible switching. This is a serious limitation if we aim to apply this to an experimental setting, where concrete knowledge of all parameters might not be feasible. Hence, the next step was to conduct a statistical analysis of this parameter-induced change in the dispersion relation.

\subsubsection*{Finding a generalized pattern switch}

To obtain results which are generalizable to more than one particular parameter set of a given network, we chose to apply the method we described in the previous section to a range of parameters. In particular, we used Latin hypercube sampling  \cite{Zheng_Shao_Ouyang_2016} to sample parameter space in a biologically relevant range. For each set of parameters sampled, we checked the conditions for pattern formation, namely that the homogeneous steady state is stable, and it becomes unstable in the presence of diffusion, and only kept the parameter sets that were able to produce patterns. This search was conducted using the code provided by Scholes et al. \cite{SCHOLES2019243}. We denote $N_s$ as the total number of sampled parameter sets, with $N_p$ and $N_{np}$ representing the sets capable and not capable of producing Turing patterns, respectively.

To generalize our findings to more complex networks than the 2-node one, we chose to compute the change in maximum of the dispersion numerically by using small perturbations to the parameters and calculating the change in $\lambda_{max}$, which we denote $\Delta \lambda_{max}$. For each of the genetic network models in Fig. \ref{fig1}A-C, we sampled a set of parameters and checked which ones could produce patterns. Once we know $N_p$ and $N_{np}$ for each model, we calculated the proportion of parameters that produce patterns, which quantifies the robustness of each model. As previously suggested by \cite{SCHOLES2019243, Vittadello_Leyshon_Schnoerr_Stumpf_2021}, we found that larger networks exhibit greater robustness within the same parameter intervals, \red{meaning that a greater fraction of sampled parameters is able to generate Turing patterns}. This can be observed in the barplots in Fig. \ref{fig2}B, which show the number of parameters satisfying the patterning conditions. For each model, we sampled $10^5$ and $10^6$ parameter sets, respectively, and observed similar results across both sample sizes. \red{We did this in order to show continuity of Turing instability in parameter space}. In the 2-node network, approximately 0.0002\% of the parameters satisfied the patterning conditions, increasing to 0.00055\% in the 3-node network and 0.0017\% in the 4-node network. From these numbers, it is clear that the increment of Turing space is remarkable, since it more than tripled from the 3-node to the 4-node network.

The ideal scenario for designing a generalized pattern switch is to identify a parameter for which the distribution of  $\Delta \lambda_{max}$ is consistently positive or negative. This would ensure that perturbing the parameter reliably changes the dispersion relation in a predictable direction, providing a robust prediction for experimental validation. For each of the parameters in the model, we obtained $N_{np}$ different values of $\Delta \lambda_{max}$. To quantify how good a set of these parameters is, we defined the following metric

\red{\begin{equation}
    E(\Omega) = \frac{|\sum_{i\in \Omega} \mathds{1}_{[0,\infty)}(i)-\mathds{1}_{(-\infty,0]}(i)|}{N_{np}},
\end{equation}}
where $\Omega$ is the set consisting of $N_{np}$ values of $\Delta k_c^2$ or $\Delta \lambda_{max}$. This function is simply the absolute value of the difference between the number of positive elements and negative elements, divided by the size of the set, so we will denote it by `sign difference'. Hence, if $E(\Omega)=1$, all elements of $\Omega$ are either positive or negative, and if $E(\Omega)=0$ then there is an equal number of positive and negative elements. Note we ranked our obtained set in descending order using this function.

For the 2- and 3-node networks, we found that the best parameters achieved a sign difference score of approximately 0.9, indicating that most of the distribution lies on one side of the $x$-axis. In contrast, the 4-node network exhibited more symmetrical distributions, with a maximal sign difference score of 0.7. The distributions for the best and worst (\red{blue and orange histograms respectively}) ranked parameters for the 2-, 3- and 4-node networks are shown in Fig. \ref{fig2}C, D, and E, respectively.  We also show the top and bottom three parameters ranked by the sign difference score for each model in Fig. \ref{fig2}F, G and H. We observe that the best score is reduced for each model as we add nodes, but the behavior of the bottom scores does not follow a clear trend. We also notice that there are several parameters that have a very high score. For the 2-node and 3-node models, shown in Fig. \ref{fig2}F and G respectively, the top four parameters have a very similar value, and then we observe a sudden drop. This behavior is better observed in the inside panels of Figs. \ref{fig2}F-H, where a lineplot of the score is showed in decreasing order. For the 4-node network, the behavior is very different, showing a much slower and constant drop in value. To assess the robustness of these results to local perturbations, we also explored small variations around individual parameter sets. As shown in Fig. S1 in SI, the change in the dispersion relation remains smooth and locally stable, with the sign of the effect preserved across a wide range of biologically plausible values.

In the 2-node network, the best switch parameters are the degradation and basal production rates of both concentrations. On the other hand, for the 3-node network we observe that the best parameters are the degradation rates of all concentrations and the basal production rate of node C. All of these parameters could be used to experimentally verify a pattern forming system, by having different values in distinct regions of space. Specifically, the 3-node model showed an asymmetry that could be explored experimentally by using different levels of the basal production rate of C, which could be distributed spatially in the culture. We will explore further possible experimental applications later on.

These results show that tuning pattern formation using the mechanism proposed in this section, i.e. shifting the maximum of the dispersion, is plausible for 2- and 3-node networks. We are capable of finding parameters, $b_B$ for the 2-node network and $\mu_A$ for the 3-node network, whose effect solely increases or decreases the dispersion, which makes them ideal parameters for tuning pattern formation.  We also saw a decrease in the suitability of parameters when we considered the 4-node network, which could imply that in networks with more nodes we cannot control the dispersion relation with single-parameter changes.

The above approach enables the design of a pattern switch, allowing precise control over the occurrence of pattern formation. In the next section, we delve into a different type of mechanism that may allow us to control the type of Turing pattern that is formed.
 
\subsection*{Pattern dial: controlling pattern type}
Next, we consider the problem of tuning the type of pattern. Most studies on pattern formation rely on linear stability analysis to determine whether patterns can form. This method, although extremely successful in the past, was previously found not completely accurate at describing pattern formation in different situations, e.g. in multistable systems \cite{Krause_Gaffney_Jewell_Klika_Walker_2024}. Furthermore, linear stability analysis is not capable of informing us about the type of pattern that the system can produce (e.g., stripes vs. spots). To determine the type of pattern the system produces, we must move beyond linear stability analysis and apply weakly nonlinear techniques,  specifically the method of multiple scales \cite{Kevorkian_Cole_1996, Nayfeh_2000}.

\begin{figure*}
\includegraphics[width=\textwidth]{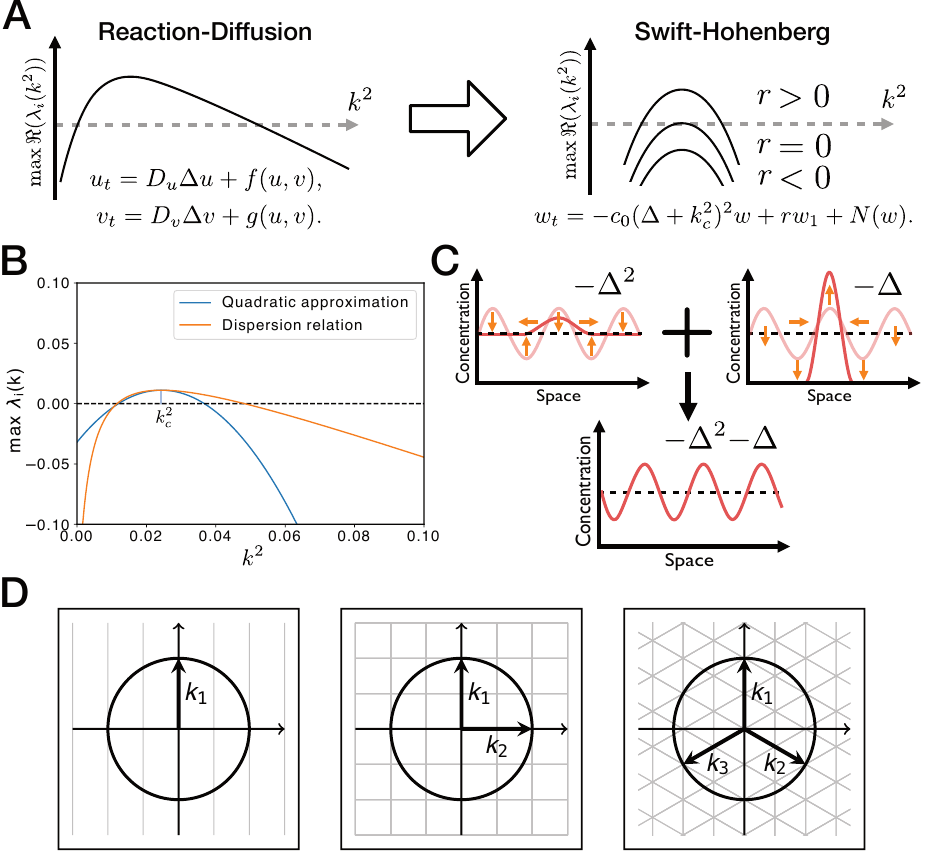}
\caption[\textbf{Illustration of the quadratic approximation and wavevector lattices.}]{\textbf{Illustration of the quadratic approximation and wavevector lattices.} \red{\textbf{(A)} Sketch of the projection from the full reaction-diffusion system to the Swift-Hohenberg model, with key parameter $r$ acting as the switch between pattern formation and no pattern.} \textbf{(B)} Application of the quadratic approximation for the dispersion relation. The original dispersion relation for a given model is shown in orange, while a quadratic fit (blue) around the maximum, $k_c^2$, leads to the Swift-Hohenberg (SH) equation. \textbf{(C)} The effect of the negative biharmonic operator and negative Laplacian operator on a cosine wave. On the left, the negative biharmonic smooths out the cosine wave, first reducing side peaks and eventually flattening the central peak to reach a homogeneous steady state. On the right, the negative Laplacian amplifies the wave's clumps, causing them to grow tighter, eventually tending toward a Dirac delta-like function in the limit as $t\rightarrow\infty$. The combination of these two opposing effects results in a stable pattern, as shown in the lower plot. \textbf{(D)} Lattices generated by varying numbers of wavevectors. A single wavevector produces stripes. With two wavevectors, a square lattice forms, supporting both stripes and squares. Three wavevectors produce a hexagonal lattice, where both hexagonal spots and stripes are stable.}
\label{fig3}
\end{figure*}

The method of multiple scales is a versatile technique widely used in fields ranging from fluid dynamics to material science. For a thorough review of this method and the different applications see \cite{Nayfeh_2000}. One of the first uses in pattern formation appeared in the study of postcritical Rayleigh-Bénard convection \cite{Newell_Whitehead_1969}. In this approach, the solution to an ordinary (ODE) or a partial (PDE) differential equation is expanded asymptotically in powers of a small parameter, typically denoted as $\epsilon$, representing the distance from the bifurcation (see Fig. \ref{fig3}A, Right, where $\epsilon$ is written as $r$). Each term in this expansion corresponds to the solution at a different scale, and the summation of these solutions gives an approximation to the full solution.

The method's key step is the emergence of solvability conditions at higher orders in the expansion. These conditions determine the parameters that control the system’s behavior and yield reduced equations that govern pattern stability. This approach was extensively applied in reaction-diffusion systems to analyze Turing patterns. For instance, it was used on a 2-dimensional reaction diffusion system to prove that stripes and spots are mutually exclusive as stable patterns, and that quadratic terms are key for spotted patterns \cite{ermentrout1991}. This result was later generalized to $n$-dimensions \cite{ndimensionalstability}. Numerous other works used this method to study stability conditions in specific 2D systems \cite{Chen_Wu_Liu_Chen_2023, Wang_Liu_2023, Reinken_Heidenreich_Bär_Klapp_2024}.

Nevertheless, application of this procedure to multi-dimensional systems is not trivial, and theoretical complications often limit the practical use of this method in experimental settings. In the next section, we present a framework to simplify multi-dimensional systems by a series of approximations, ultimately mapping them to a one-dimensional model where applying the method of multiple scales is more tractable. We then explore how well these approximations hold and whether the results from the 1D model apply to the original multi-dimensional system.

\subsection*{Mapping to the Swift-Hohenberg equation}

Our starting point for this section is a general reaction-diffusion system where the nonlinearities are not yet specified. This type of system can describe a wide range of phenomena, including gene interactions. These models are particularly relevant for the study of synthetic Turing patterns in bacterial colonies. Here, we show how a general reaction-diffusion model can be reduced to a Swift-Hohenberg-like equation \red{(details and derivation in the SI)}. Later on, we apply this reduction to a specific case where the nonlinear functions are given by Hill equations, as in the model from Eq. \ref{network2d}.

We begin with a general two-dimensional reaction-diffusion equation:
\begin{equation} \label{RDGeneralch4} \begin{split} \partialder{u} = D_u \Delta u + f(u,v),\\\partialder{v} = D_v \Delta v + g(u,v), \end{split} \end{equation}where $u$ and $v$ are two-dimensional functions representing concentrations of two interacting substances, and $f$ and $g$ are general nonlinear functions.

To simplify the system, we first linearize the equations around a fixed point, $(u_0, v_0)$, which allows us to focus on the dynamics near equilibrium. We do this by performing a Taylor expansion of the nonlinear terms, which captures the behavior of small perturbations:

\begin{equation}\label{RDGen_exp}
\begin{split}
    \partialder{u} = D_u \Delta u +f_u u+ f_v v + \mathcal{HOT}_u, \\
    \partialder{v} = D_v \Delta v +g_u u + g_v v +\mathcal{HOT}_v,
\end{split}
\end{equation} 
where $f_u=\partial f/\partial u$ is the partial derivative of $f$ with respect to $u$ evaluated at $(u_0,v_0)$ and similarly for the other terms, and $\mathcal{HOT}_u$ and $\mathcal{HOT}_v$ are the higher order terms for $u$ and $v$, respectively. To avoid unnecessary terms, we redefine the variables $u$ and $v$ to be centered around $(u_0,v_0)$. The first approximation to the original model involves truncating the Taylor expansion in Eq. \ref{RDGen_exp} after the third-order terms. This is referred to as the Taylor-approximated (TA) model. 

 Next, we take the Fourier transform of this TA model and apply a transformation to diagonalize the Jacobian $\mathbf{J}_k$. By assuming there is a single Turing instability which only occurs for one of the eigenvalues, we can approximate the dispersion relation with a quadratic function, which, by returning to the spatial domain with an inverse Fourier transform, leads to the following model:

 \begin{equation}
\label{RDnewmodel}
\begin{split}
    \partialder{w_1} &= -c_0 (\Delta+k_c^2)^2 w_1 +r w_1 +\mathcal{HOT}_{w_1},\\
    \partialder{w_2}& = \gamma \Delta^2 w_2 + b w_2 +\mathcal{HOT}_{w_2}.
\end{split}
\end{equation}

This system, which we denote as the quadratic-diagonal (QD) model, is the second step in our approximation to a one-dimensional model. The full derivation of this model, together with an explanation on the terms $\mathcal{HOT}_{w_1}$ and $\mathcal{HOT}_{w_2}$, can be found in SI. 

The first equation of the QD model is a Swift-Hohenberg-like model, with a biharmonic and a diffusion term. The second equation represents simple diffusion. This type of model was previously considered in motility-induced phase separation (MIPS) \cite{Martínez-Calvo_Wingreen_Datta_2023}. Next, we reduce this equation to a single Swift-Hohenberg-type equation. By considering $w_2$, we can see that the only contribution to linear order is $b w_2$, where $b$ is the value of $\lambda_2$ when $k=0$. Since the initial state at $k=0$ must be stable, $b<0$, and because up to linear terms, $w_1$ and $w_2$ are independent, then $w_2$ decreases exponentially. Hence, for small times, we expect $w_2$ to be very small, allowing us to set $w_2=0$. After applying these approximations, we derive a Swift-Hohenberg equation for $w_1$:

\begin{equation}
\label{Swift-Hohenberg}
    \partialder{w_1} = -c_0 (\Delta+k_c^2)^2 w_1 +r w_1 +N(w_1),
\end{equation}
where the nonlinearity takes the form $N(w_1) = c_1 w_1^2 + c_2 w_1^3$. This is similar to fitting a quadratic polynomial to the dispersion relation, as showcased in Fig. \ref{fig3}B. Although this reduction simplifies the system, it retains the key dynamics necessary to model Turing patterns.

\subsection*{Application of SH projection and amplitude equation}

We are now in a position to apply the SH model projection and amplitude-equation formalism to parameter values that generate Turing patterns in the 2-node network from Eq. \ref{network2d}. For a more detailed introduction to the amplitude equation, see the SI. We use the amplitude equation derived for three different wavevectors \cite{Cross_Greenside_2009, Hoyle_2006}, which describes the evolution of the pattern amplitudes $A_1$, $A_2$, and $A_3$ over time:

\begin{equation}
\label{amplitudeeq_3wvCh4}
\begin{split}
\frac{d A_1}{dT} = r A_1 + 2 c_1 \bar{A_2}\bar{A_3} - \gamma_1  A_1 |A_1|^2 - \beta_1 A_1 (|A_2|^2+|A_3|^2), \\
\frac{d A_2}{dT} = r A_2 + 2 c_1 \bar{A_1} \bar{A_3}- \gamma_2 A_2 |A_2|^2 - \beta_2 A_2 (|A_1|^2+|A_3|^2), \\
\frac{d A_3}{dT} = r A_3 + 2 c_1 \bar{A_1} \bar{A_2}- \gamma_3 A_3 |A_3|^2 - \beta_3 A_3 (|A_1|^2+|A_2|^2),
\end{split}
\end{equation}
where $\gamma_1 =\gamma_2=\gamma_3 = 3 c_2 - \frac{38 c_1^2}{9 k_c^4 c_0} $ and $\beta_1=\beta_2=\beta_3 = 6 c_2 - \frac{5 c_1^2}{k_c^4 c_0}$. \red{The number of wavevectors will determine which types of patterns we will be able to distinguish. In this case, three is enough to distinguish spots and stripes, as shown in Fig. \ref{fig3}D.} The values of these parameters will determine the stability of the different pattern types, as we shall see later. 

We considered 5 different parameter sets from the original model, which are shown in the S1 parameter table in SI. When considering the projection into the SH model, we observed two different types of behaviour. Three of the parameter sets produced an amplitude equation with $\gamma_1<0$. We call this parameter set `non-saturating' because the cubic term is insufficient to saturate the amplitude equation \cite{Hilali_1995}, and as such, it will not be able to describe patterns unless we perform an expansion up to higher-order terms. We focus instead on two other parameter sets that do saturate the amplitude equation, referred to as 'saturating' parameter sets.

\begin{figure*}
\includegraphics[width=\textwidth]{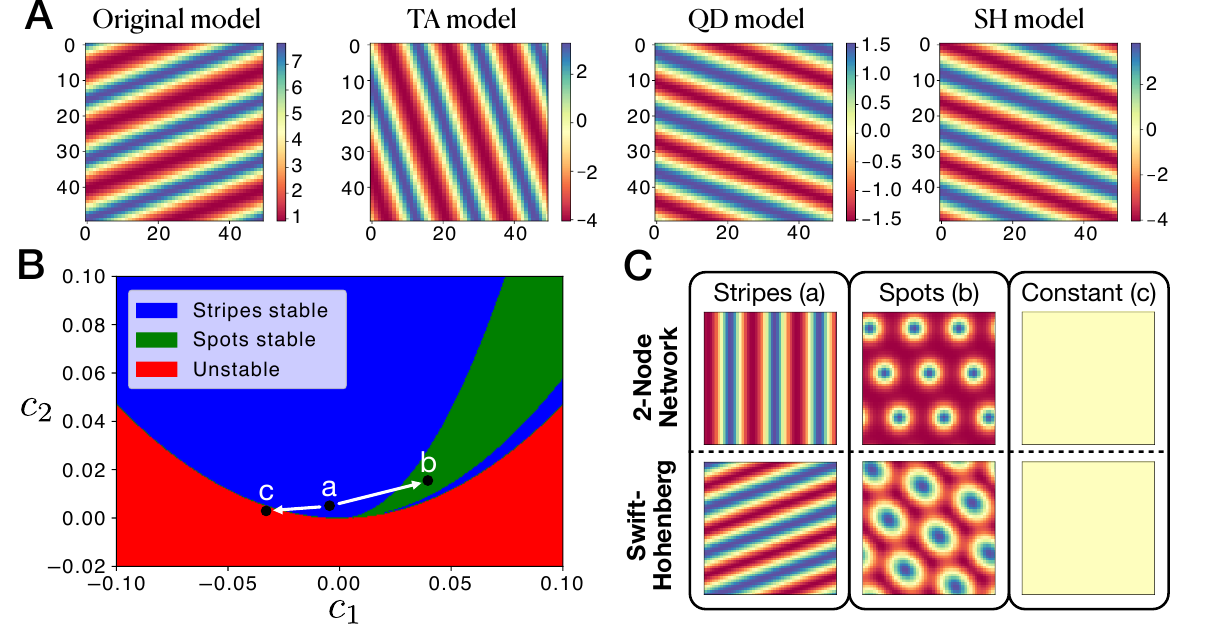}
\caption[\textbf{Model approximations, stability regions and pattern transition.}]{\red{\textbf{Model approximations, stability regions and pattern transition.} \textbf{(A)} Different patterns produced by a saturating parameter set (the last in Table S1 in the SI), where the cubic nonlinearity is sufficient to saturate the amplitude equation. We show from left to right the original model, the Taylor-approximated (TA) model, the quadratic-diagonal (QD) model and the Swift-Hohenberg (SH) model. All approximations produce the same striped pattern for this parameter set. Orientation of stripes changes depending on initial conditions. \textbf{(B)} Stability diagram in $c_1-c_2$ space, obtained from the amplitude equation. The original parameters correspond to a projection of saturated parameters from Eq. \ref{network2d} to the Swift-Hohenberg equation. The different colors show the different areas where each pattern is stable.  Blue corresponds to regions where only stripes are stable, green to regions where only spots are stable and red corresponds to unstable regions. Originally our system was at point \textbf{a}, where only stripes are stable. Investigating the effect of varying different parameter by using the projection, we found that increasing $k_A$ leads to the spots stable region (point \textbf{b}), whereas decreasing it leads to the unstable region (point \textbf{c}). \textbf{(C)} Simulations of the different parameter sets at points \textbf{a}, \textbf{b} and \textbf{c} for the 2-node network (above) and the Swift-Hohenberg model (below), resulting in the validation of the analytical prediction of pattern change.}}
\label{fig4}
\end{figure*}

With these two parameter sets, we observed similar results across all models\red{, as shown in Fig. \ref{fig4}A.} Using periodic boundary conditions allowed the patterns to develop their most stable wavelength, regardless of domain size. We observed that all models produced stripes of similar wavelengths, even though orientation was driven by the initial conditions. If a different set of boundary conditions was used, in this case the Neumann boundary conditions, the pattern for the initial model seemed to shift into a checkerboard pattern, which was not reproducible in the other models. However, this is thought to be an artefact caused by the effect of the Neumann boundary conditions.

Next, we apply the stability conditions from the amplitude equation. First,  starting from the saturating parameter values we obtain a projection onto the Swift-Hohenberg equation. In particular, we obtain the following parameters: $c_0 = 10.53748, c_1 = -0.00449, 
c_2 = -0.00508$, and $r = 0.05993$, with the maximum wavenumber at $k_c^2 = 0.16866$. Applying these values to the conditions obtained from the amplitude equation (Eq. \ref{amplitudeeq_3wvCh4}), we obtain the stability diagram in Fig. \ref{fig4}B. The diagram in the $c_1-c_2$ plane is divided into three regions. Blue indicates stable stripe patterns, green indicates stable dot patterns, and red indicates instability in both patterns. We could also have regions where both patterns are stable, and obtaining one or the other would depend on the initial conditions, \red{but these were not present in the parameter range explored}. The parameter set used to obtain this diagram is marked by point (a), which is in the region where only stripes are stable. We can observe that the most frequent regions contain either stable stripes or no stable patterns. 

To observe transitions between pattern types, we need to move into the green region, which corresponds to an increase in parameter $c_1$. By modifying the original parameter set and calculating the projection onto the Swift-Hohenberg model, we found that the parameter $k_A$ was capable of transitioning towards the dot-stable region. By simulating the parameter values at different points along this trajectory, we confirm that the predictions from the amplitude equation are indeed fulfilled, \red{as shown in Fig. \ref{fig4}C}. As we increase $k_A$, we can see a transition to spots of the generated pattern in the 2-node network (see point (b), top pattern). We can also see this transition if we take the $c_1-c_2$ values from the stability diagram and simulate the Swift-Hohenberg equation using those values while keeping the rest of the parameters fixed. Similarly, if we decreasing the $k_A$ value, we are able to transition into the unstable region, where both the 2-node network and the Swift-Hohenberg models stop exhibiting pattern formation. 




\subsection*{Application to an spatially distributed parameter}

Finally, we investigated the application of both of these switches by using a spatial distribution in some of the parameters. In particular, we can rewrite the 2-node network with spatially dependent parameters as

 \begin{equation}\label{network2dspdist}
\begin{split}
    \partialder{A} &= b_A(\mathbf{x})+V_A(\mathbf{x})\frac{1}{1+(k_A(\mathbf{x})/A)^2}\frac{1}{1+(B/k_{BA}(\mathbf{x}))^2}-\mu_A(\mathbf{x}) A+D_A(\mathbf{x}) \Delta A ,\\
    \partialder{B} &= b_B(\mathbf{x})+V_B(\mathbf{x})\frac{1}{1+(k_{AB}(\mathbf{x})/A)^2}-\mu_B(\mathbf{x}) B+D_B(\mathbf{x}) \Delta B.
\end{split}
\end{equation}

This equation contains a spatial dependence for all the parameters, but in practice, we focus on a given parameter that serves as a switch. We first consider the pattern switch. We saw before that $\mu_A$ is capable of acting as a pattern switch: a high $\mu_A$ was capable of increasing the dispersion relation and a low $\mu_A$ decreased it. If we choose two values, $\mu_{A}^{on}$ and $\mu_{A}^{off}$, for which the dispersion is above and below zero respectively, then we can define a spatial distribution for $\mu_A(\mathbf{x})$ in terms of these values. In particular, assuming a square domain  $[0,100]\times[0,100]$, we can define:

\begin{equation}
\mu_A(\mathbf{x}) =
\begin{cases} \mu_{A}^{on}
 & \text{if $y > 50$}, \\
\mu_{A}^{off} & \text{else}.
\end{cases}
\end{equation}

\begin{figure*}[t!]
\includegraphics[width=\textwidth]{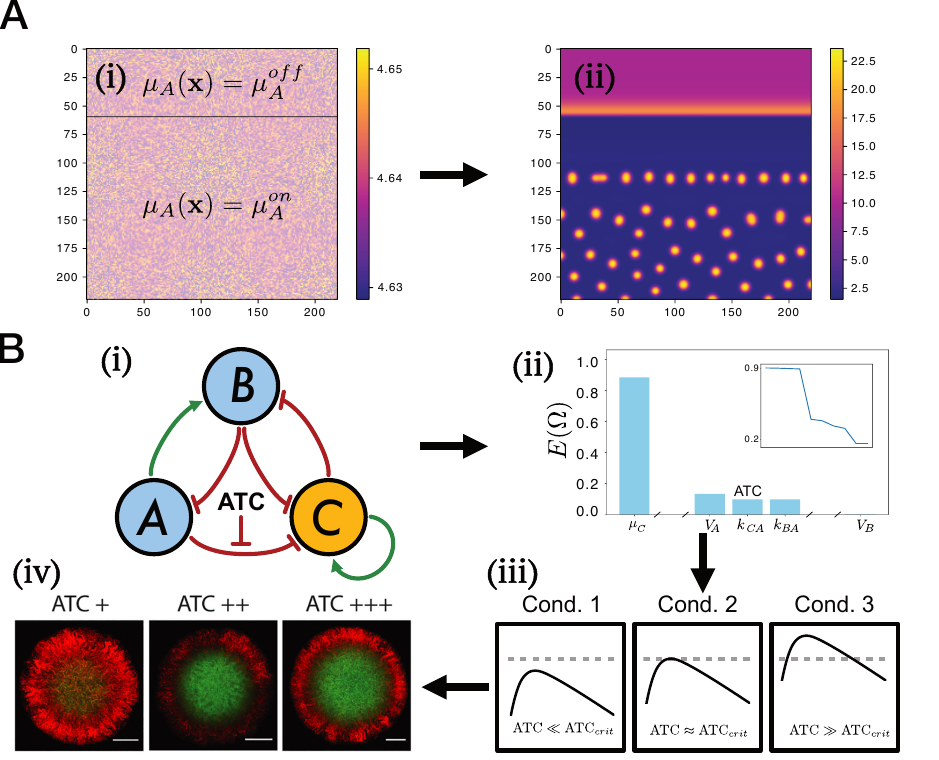}
\caption[\textbf{Potential experimental applications of the pattern switch.}]{\textbf{Potential experimental applications of the pattern switch.} \textbf{(A)} An initial spatial distribution of the parameter $\mu_A$ allows the appearance of pattern formation in specific regions of space. (i) We let $\mu_A = \mu_A^{off}$, a value for which the dispersion is below zero, for the upper region of space, and $\mu_A = \mu_A^{on}$, a value which allows for pattern formation, for the lower region. (ii) We simulate the dynamics, allowing this parameter of the PDE to have the different spatial values and observe that pattern formation indeed is only produced in the region of space where $\mu_A = \mu_A^{on}$. \textbf{(B)} Workflow for using the pattern switch to test for different levels of concentrations of a parameter. (i) A genetic network where anhydrotetracycline  (ATC) is used to suppress the inhibition from node $A$ to $C$ ($k_{CA}$ in the 3-node network introduced before). (ii) Checking the effect of ATC. As it only has a moderate effect, it is not the best pattern switch (although it still can be used). (iii) Determining three different conditions for which ATC can act as a pattern switch. (iv) Experimentally test of those conditions. Images in (iv) were taken from Fig. 1A in \cite{martina2023} with permission. Note the concentration of ATC was not computed using our workflow. Nevertheless, the data show some similarity to our approach. }
\label{fig5}
\end{figure*}

Using this parameter with spatial dependence, we should be able to devise where in parameter space we observe the pattern. An example simulation of this is shown in Fig. \ref{fig5}A, where the parameter $\mu_A$ has a value of $\mu_{A}^{off}$ in the upper part of the spatial grid, and a value of $\mu_{A}^{on}$ in the remainder. The initial condition is normally distributed noise around the fixed point. If we numerically solve the system, then we obtain Fig. \ref{fig5}B. In this graph we can clearly see that the region of parameter space related to $\mu_{A}^{off}$ appears to reach a steady state, while the region of parameter space with $\mu_{A}^{on}$ indicates pattern formation. More complicated geometries are possible, but constraints due to the boundary conditions might destroy the pattern. 

Once a parameter is determined for strong switch-like behavior, it is also beneficial to experiment with different conditions on the parameter values. This is demonstrated in Fig. \ref{fig5}B(i-iv), where anhydrotetracycline (ATC) is employed to slow down the degradation of node A to node C, which we verified to be a relatively good parameter switch.

\section*{Discussion}

We introduced two complementary strategies for controlling Turing pattern formation in synthetic biological systems: the pattern switch and the pattern dial. These approaches enable both coarse (on/off) and fine-grained (pattern type) control, offering practical tools for designing and engineering spatial organization in biological contexts.

The pattern switch provides a simple yet powerful mechanism for toggling Turing patterns by adjusting a single parameter that alters the system’s dispersion relation (Fig.~\ref{fig1}D). While this idea is well-established in theoretical models, we extended it to realistic genetic networks and developed a systematic method to identify which parameters act as effective switches. By computing a sign difference metric, we ranked parameters by their ability to shift the system between stable and unstable regimes. This analysis revealed that small networks are especially amenable to control, where a handful of key parameters can strongly influence pattern formation (Fig.~\ref{fig2}F, G). In contrast, larger networks exhibit greater robustness to parameter changes but reduced controllability (Fig.~\ref{fig2}H), highlighting a fundamental trade-off between tunability and robustness. This mirrors previous findings \cite{SCHOLES2019243} and may reflect evolutionary design principles in natural systems where patterns—for example, for camouflage—must be both evolvable and resilient \cite{Kirschner1998Evolvability}.

Our switch-based framework has clear experimental potential, particularly because it simplifies the challenge of working with poorly characterized parameters. Key rates, such as degradation or basal expression (e.g., $\mu_A$, $b_B$), can be tuned or spatially modulated to induce localized pattern formation (Fig.~\ref{fig5}A), without requiring full parameter identifiability. This makes the pattern switch a practical entry point for synthetic pattern engineering.
In contrast, the pattern dial offers a more refined control mechanism that allows continuous tuning of pattern morphology—for example, transitions from stripes to spots (Fig.~\ref{fig1}E). By projecting the genetic network onto the Swift–Hohenberg equation and applying weakly nonlinear analysis, we derived amplitude equations that predict stable pattern types. This analytical framework, building on earlier work \cite{Hiscock_Megason_2015}, captures features invisible to linear stability analysis alone \cite{Cross_Greenside_2009} and identifies conditions under which pattern transitions occur (Fig.~\ref{fig4}B,C). While accurate near the Turing bifurcation, the approach becomes less reliable farther away, where higher-order nonlinearities play a greater role \cite{Hilali_1995}. Despite these limitations, the qualitative trends remained consistent with predictions, underscoring the framework’s utility.

The experimental implementation of the pattern dial is more challenging, as it requires precise parameter knowledge and perturbations within the weakly nonlinear regime\red{, unlike the pattern-switch, which is generalizable to all parameters in a given network structure}. Nevertheless, our identification of a candidate dial parameter ($k_A$ in the 2-node network) provides a promising target. In the future, parameter inference tools such as PINNs \cite{TuringPinns, Matas-Gil_Endres_2024} or symbolic regression \cite{schnorrimperial} could help bring the dial closer to practical use. Together, the switch and dial form a complementary toolkit: the switch offers accessible, spatially localized control with minimal parameter assumptions, while the dial enables the engineering of desired geometries, albeit with tighter experimental constraints. These approaches can be extended further—for example, combining multiple switch parameters for nonlinear control, or including higher-order corrections to extend the dial’s predictive regime.

Beyond individual control mechanisms, our work sheds light on how network architecture shapes controllability and robustness. Smaller networks allow precise, switch-like control, while larger networks are harder to steer but naturally more stable—suggesting a design tension that may influence both synthetic and natural patterning systems. These strategies integrate well with existing control paradigms for pattern formation, such as logic gate-based dose–response filters \cite{CBarnes_2024}, and multicellular coordination of gene expression \cite{Salzano_Shannon_Grierson_Marucci_Savery_DiBernardo_2023}. We envision that combining binary, analogue, and spatial controls will form the foundation of a control theory of biological pattern formation, as envisioned in recent work \cite{DelVecchio_Qian_Murray_Sontag_2018}.

In conclusion, the pattern switch and pattern dial represent experimentally actionable and theoretically grounded strategies for controlling Turing patterns. By offering scalable and flexible means to design spatial organization, they bridge the gap between mathematical theory and synthetic morphogenesis, laying the groundwork for future applications in synthetic tissues and programmable patterning.

\section*{Methods}
\subsection*{Turing conditions}

In the following, we provide a brief account of the conditions for the diffusion-driven Turing instability and onset of pattern formation which underlie most of the mechanisms proposed in this paper \cite{Murray2}. We start by linearizing the 2-dimensional model in Eq. \ref{RDGeneralch4}, which yields 

\begin{equation}
\mathbf{u}_t = 
\begin{pmatrix}
f_u  & f_v \\
g_u  & g_v 
\end{pmatrix} 
\mathbf{u} +D \Delta \mathbf{u}
 = \mathbf{J u}+ D \Delta \mathbf{u},
\end{equation}
where $D= \text{diag}(D_1,D_2)$ is the diagonal matrix of diffusion coefficients and $\mathbf{J}$ is the Jacobian of the nonlinearity functions evaluated at $(u_s,v_s)$, which is assumed to be a homogeneous fixed point. 
The eigenvalues of the Jacobian determine the stability of the steady state. We assume $(u_s,v_s)$ is an asymptotically stable fixed point, that is, the real part of both eigenvalues is negative without diffusion. To assess the impact of diffusion, and verify the existence of a diffusion-driven instability, we Fourier transform our equation into
\begin{equation}
\mathbf{J}\tilde{\mathbf{u}} - D k^2 \tilde{\mathbf{u}} = \mathbf{J_k}\tilde{\mathbf{u}}, \quad
\mathbf{J_k} = 
\begin{pmatrix}
f_u - D_u k^2 & f_v \\
g_u & g_v - D_v k^2
\end{pmatrix}.
\end{equation}

This yields a $k$-dependent Jacobian, which means that the stability of the system will depend on the wavenumber $k=|\mathbf{k}|$ of the perturbations. We can characterize the stability by looking at the maximum of the real part of all eigenvalues, or $\max(\Re(\lambda_i))$: if it is positive, the system is unstable, but if it is negative, the system is stable. As such, we will focus on the behavior of $\max(\Re(\lambda_i))$ as a function of $k^2$. This function, which we may write as $\max(\lambda_i(k))$, is denoted as the dispersion relation. For a classic Turing instability the system is stable for small $k^2$, but becomes unstable for a finite range of $k^2$ values. This means that perturbations following a given wavelength will dominate, causing a spatial pattern with a defined wavelength to appear. For $\max(\lambda_i(k))$, this means that at $k^2=0$ we must have $\max(\lambda_i(0))<0$, and that for some values $k_1^2$ and $k_2^2$, we must have $\max(\lambda_i(k))>0$ only for $k_1^2<k^2<k_2^2$. This is represented in Fig. \ref{fig3}A. Obtaining conditions that satisfy these requirements leads to numerical conditions for pattern formation.

\subsection*{Numerical simulations}

We numerically solved the various reaction-diffusion PDEs and obtained Turing pattern solutions shown in the different figures. These were generally performed on a square grid of $50\times50$, with periodic boundary conditions. An exception to these boundary conditions is the result shown in Fig. \ref{fig5} AII, where we imposed periodic boundary conditions along the $Y$-axis and Neumann boundary conditions along the $X$-axis, to avoid breaking the pattern simply by the effect of the boundary conditions. \red{This behaviour was due to spots only emerging far from the fixed boundaries. Instead of elongating the $x$-axis, due to computational limitations, we chose to use mixed boundary conditions.}

Initial conditions were, as usual for Turing patterns, normally distributed noise around the homogeneous steady-state solution of the PDE. We then used a spatial and temporal discretisation to numerically integrate the system forward in time using the Explicit Runge-Kutta method of order 5 (RK45), implemented in \textit{SciPy}. For the spatially distributed parameter in Fig. \ref{fig5}A, we used an alternating-direction-implicit Crank-Nicolson method, whose efficiency allowed for a higher spatial resolution ($100\times100$ grid).

\subsection*{Numerical methods of the pattern switch}

To perform the parameter search and find the different steady states capable of producing Turing patterns we used a similar procedure to the one taken in \cite{SCHOLES2019243}. We used Latin hypercube sampling to explore parameters in a biologically relevant region, and then for each parameter, calculated the steady states (there can be more than one). In contrast to \cite{SCHOLES2019243}, in which dynamics with different starting points were calculated to determine sinks (attractors), we used a modification of the Powell hybrid method to find the roots of a vector function, with different random initializations. This implementation is available in \textit{SciPy} under the \textit{root} function. These steady states were then classified according to Turing's conditions as Turing or no Turing.

Once the Turing parameter sets were obtained, for each of them we performed a sensitivity analysis as described previously. We calculated two dispersion relations: the first being the dispersion relation with the initial parameters and the second the dispersion relation after the parameter change. Once we know both dispersion relations we can compute the difference in the peak between both of them, and this is what we called $\Delta \lambda_{max}$. We repeated this for all parameters and all parameter sets to obtain the distributions shown in Fig. \ref{fig2}C-E.

\section*{Acknowledgements} We thank Mark Isalan and his lab members for valuable discussions and explanations of their experiments. This research was funded through a studentship (to AMG) from the Department of Life Sciences at Imperial College London, which again is supported by the Biotechnology and Biological Sciences Research Council (BBSRC) and other funding sources. The work was further supported by a UKRI Transition Award: Artificial Intelligence for Engineering Biology (BB/W013770/1) (to RGE).

\section*{Declaration of Interests}
The authors declare no competing interests.

\bibliographystyle{vancouver}
\bibliography{refs.bib}
\end{document}